# Astrochemical Evolution Step From Acenaphthylene C12H8 To Pure Carbon C12 Around A Herbig Ae Young Star

NORIO OTA

Graduate School of Pure and Applied Sciences, University of Tsukuba,
1-1-1 Tenodai Tsukuba-city 305-8571, Japan;   n-otajitaku@nifty.com

Astrochemical evolution step of polycyclic aromatic hydrocarbon (PAH) around a Herbig Ae young star was analyzed using the first principles quantum chemical calculation. For simplicity, model molecule was selected to be acenaphthylene ($C_{12}H_8$) with hydrocarbon one pentagon combined with two hexagons. In a protoplanetary disk, molecules are illuminated by high energy photon from the central star and ionized to be cation $(C_{12}H_8)^{n+}$. Calculation shows that from n=0 to 6, molecules keeps its polycyclic hydrocarbon configuration. Whereas, at ionization step n=7, there occurs dehydrogenation of $(C_{12}H_8)$ to pure carbon $(C_{12})$. Such polycyclic pure carbon (PPC) would be attacked again by photons. At a stage of eighth ionization $(C_{12})^{8+}$, polycyclic configuration was decomposed to aliphatic carbon chains, C9, C2, and mono carbon C1. Infrared spectra (IR) of those steps were calculated to identify observed one. Carrier molecules of Herbig Ae star WW Vul (also HD145263) were identified by a combination of $(C_{12}H_8)^{2+}$ and $(C_{12}H_8)^{1+}$. Complex spectrum of HD37357 could be explained by a mixture of $(C_{12}H_8)^{2+}$, $(C_{12}H_8)^{3+}$, and $(C_{12}H_8)^{1+}$. Pure carbon molecules play an important role. IR of HD37258 was analyzed by a mixture of pure carbon $(C_{12})^{2+}$, hydrocarbon $(C_{12}H_8)^{2+}$ and $(C_{12}H_8)^{0+}$. Also, IR of HD38120 could be reproduced by $(C_{12})^{2+}$, $(C_{12}H_8)^{2+}$ and $(C_{12}H_8)^{3+}$. Those acenaphthylene related molecules are just a typical example. We should apply various size molecules to understand total view around a young star.

## 1, INTRODUCTION

In order to understand astrochemical evolution step around a Herbig Ae young star, identification of carrier molecules of infrared spectra ( IR) is very important.  Herbig Ae stars are pre-main-sequence stars having few solar masses. If we could identify hidden polycyclic aromatic hydrocarbon (PAH) carrier molecules, we could suggest one inevitable rout of a building block of "life" in the universe. There is an important observed IR data edited by B. Acke et al. including 53 Herbig Ae stars (Acke 2010). In our previous papers, we could identify those observed spectra by coronene ($C_{24}H_{12}$) and acenaphthylene ($C_{12}H_8$) related molecules based on the first principles quantum chemical analysis (Ota 2014, 2015, 2017a, 2017b, 2017c, 2017d, 2017e, 2018). In this paper, detailed astrochemical evolution step will be studied considering deeper photoionization process due to high energy photon irradiation by the central star. Here, for simplicity, ionized acenaphthylene $(C_{12}H_8)^{n+}$ molecules will be analyzed, which has a simple structure with hydrocarbon one pentagon combined with two hexagons. Every infrared spectrum and atomic configuration of $(C_{12}H_8)^{n+}$ should be calculated. We can expect dehydrogenation at deeper ionized step. Dehydrogenated polycyclic pure carbon (PPC) will be photo-ionized again until decomposition from polycyclic to aliphatic chain.

## 2, ASTROCHEMICAL EVOLUTION MODEL

By tracing recent studies unveiling astrophysics of just born young star like Herbig Ae star, we can imagine astrochemical evolution step of PAH's. Here, we like to show one typical example in case of acenaphthylene ($C_{12}H_8$). Figure 1 show a model of evolution step.

(1)  Pure carbon graphene like molecule ($C_{13}$) will be created by dust ejection after supernova explosion ( Nozawa 2003, 2006 ).

(2)  Such pure carbon molecules enter into mother dust cloud and will be modified by previously ejected hydrogen. Hydrogenated molecule ($C_{13}H_9$) has hydrocarbon three hexagons.

(3)  When new born star is creating its planetary system, protoplanetary disk may include ($C_{13}H_9$) in its stable core.

(4)  At front surface of protoplanetary disk, high speed proton from the central star may attack ($C_{13}H_9$) and kick out carbon and hydrogen. Original molecule may transform to void induced PAH to be acenaphthylene ($C_{12}H_8$).

(5)  High energy photon from the central star will illuminate such void-PAH and kick out electrons to make cationic molecules ($C_{12}H_8)^{n+}$. We should study very deep ionized case as like n=5, 6 and more. It is unusual on the earth, but reasonable in space because of ultralow molecular density (1~100 /cm³ compared with $10^{23}$/cm³ on the earth).

(6)  Continued photoionization may bring dehydrogenation to be polycyclic pure carbon PPC ($C_{12})^{n+}$.

(7)  Final step of photoionization would transform PPC to aliphatic carbon chain.

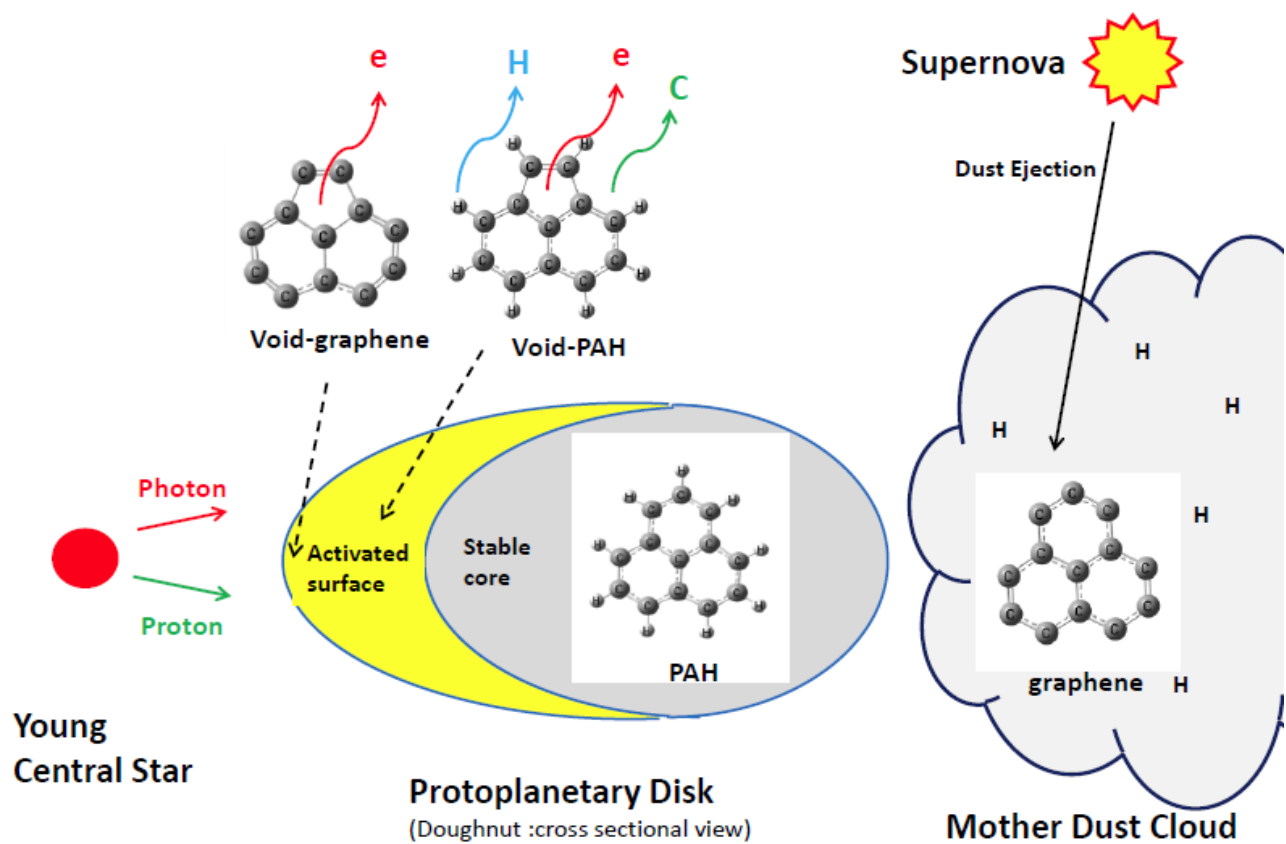

Figure 1, Astrochemical evolution model in case of acenaphthylene related molecules. Polycyclic pure carbon ($C_{13}$) will be ejected by supernova into mother dust cloud. There occurs hydrogenation to be ($C_{13}H_9$) by previously ejected hydrogen. At activated surface of protoplanetary disk, ($C_{13}H_9$) will be transformed to void induced molecule ($C_{12}H_8$) by serious attack of high speed proton from central star. Void-PAH ($C_{12}H_8$) will be ionized by high energy photon to be ($C_{12}H_8)^{n+}$. Deep photoionization may dehydrogenase hydrocarbon to pure carbon ($C_{12}$). High energy photon again attacks to make cation ($C_{12})^{n+}$ and finally brings destruction from polycyclic configuration to aliphatic carbon chain.

## 3, CALCULATION METHOD

In  quantum chemistry calculation, we have to obtain total energy, optimized atom configuration, and infrared vibrational mode frequency and strength depend on a given initial atomic configuration, charge and spin state Sz. Density functional theory (DFT) with unrestricted B3LYP functional was applied utilizing Gaussian09 package (Frisch et al. 1984, 2009) employing an atomic orbital 6-31G basis set. The first step calculation is to obtain the self-consistent energy, optimized atomic configuration and spin density. Required convergence on the root mean square density matrix was less than $10^{-8}$ within 128 cycles. Based on such optimized results, harmonic vibrational frequency and strength was calculated. Vibration strength is obtained as molar absorption coefficient ε (km/mol.). Comparing DFT harmonic wavenumber $N_{DFT}$ (cm⁻¹) with experimental data, a single scale factor 0.965 was used (Ota 2015).

Concerning a redshift for the anharmonic correction, in this paper we did not apply any correction to avoid over estimation in a wide wavelength representation from 2 to 30 micrometer.

Corrected wave number N is obtained simply by N (cm$^{-1}$) = N$_{DFT}$ (cm$^{-1}$) x 0.965.

Wavelength $\lambda$ is obtained by $\lambda$ (micrometer) = 10000/N(cm$^{-1}$).

Reproduced IR spectrum was illustrated in a figure by a decomposed Gaussian profile with full width at half maximum FWHM=4cm$^{-1}$.

## 4, PHOTOIONIZATION of $(C_{12}H_8)$ and $(C_{12})$

In Figure2, calculated molecular configuration and stable energy were illustrated.

(1) Photoionization of $(C_{12}H_8)$: Starting molecule is neutral acenaphthylene $(C_{12}H_8)$ having hydrocarbon one pentagon combined with two hexagons. Molecular stable energy was defined to be $E_0(C12H8)=0.0eV$. Photoionized state E1 of mono-cation $(C_{12}H_8)^{1+}$ is 7.5eV. Basic configuration does not change until $(C_{12}H_8)^{6+}$ with $E_6=122.1eV$. There occurs drastic change at n=7. All of hydrogen was removed from carbon network as shown in a right side blue column. Charge of every removed hydrogen were +0.5~0.7e. Remained carbon may have neutral $(C_{12})^{0+}$ or mono cation $(C_{12})^{1+}$.

(2) Photoionization of $(C_{12})$: After very deep photoionization of $(C_{12}H_8)^{n+}$, remained carbon skeleton $(C_{12})$ would be attacked by high energy photon many times. Molecular energy again defined as $E_0(C12)=0.0eV$. Until to seventh ionized state $(C_{12})^{7+}$, configuration was kept consisting pure carbon one pentagon combined with two hexagons. There occurs sudden change at ionized state n=8, as shown in green column, polycyclic $C_{12}$ was destructed to carbon chains $C_8$, $C_2$, and mono carbon $C_1$.

Calculated infrared spectra from $(C_{12}H_8)^{0+}$ to $(C_{12}H_8)^{7+}$ were illustrated in Figure 3, also from $(C_{12})^{0+}$ to $(C_{12})^{7+}$ were in Figure 4..

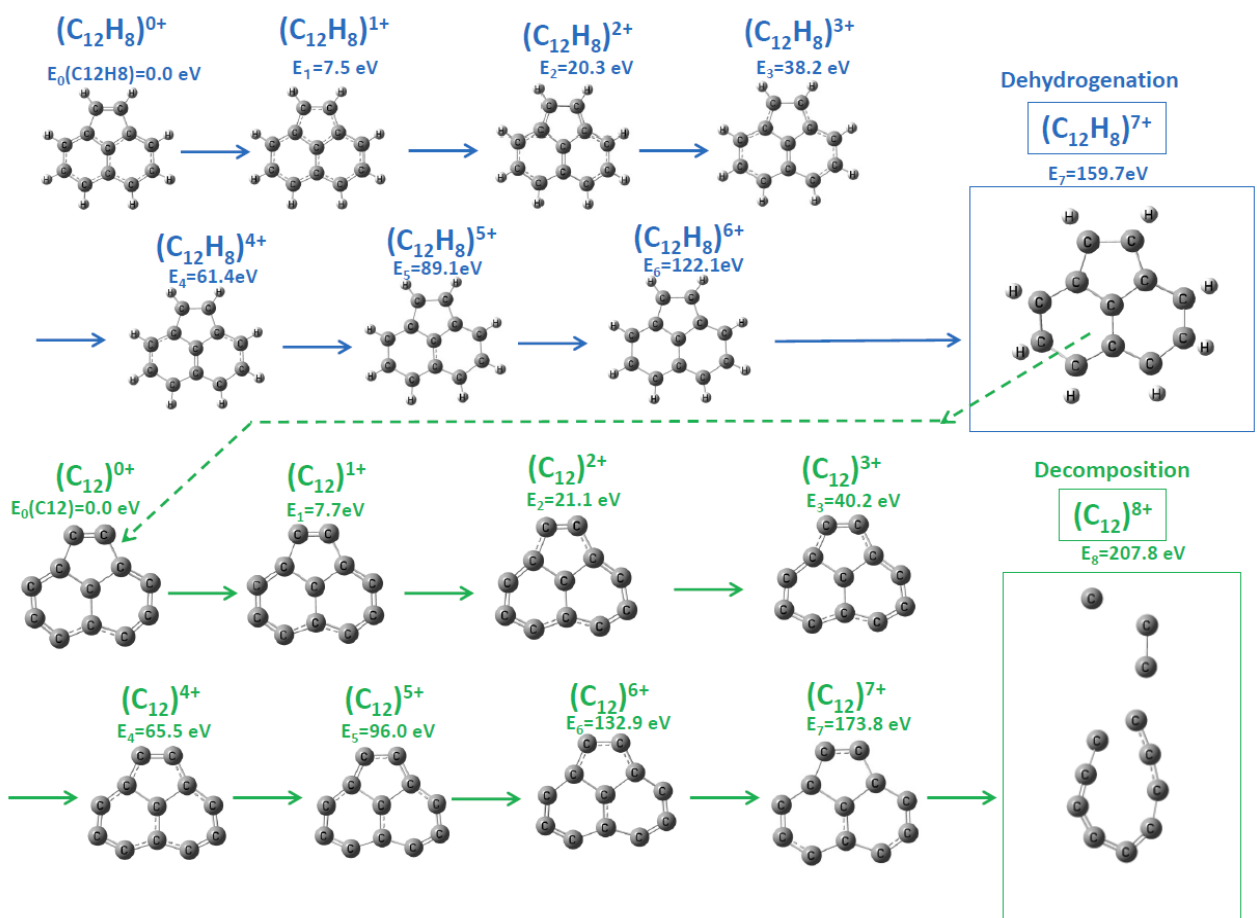

Figure 2, Photoionization of $(C_{12}H_8)$ and $(C_{12})$. At ionized state of $(C_{12}H_8)^{7+}$, there occurs dehydrogenation to polycyclic $(C_{12})$. Remained $(C_{12})$ will be attacked again by high energy photon and destructed to carbon chains $C_8$, $C_2$ and $C_1$.

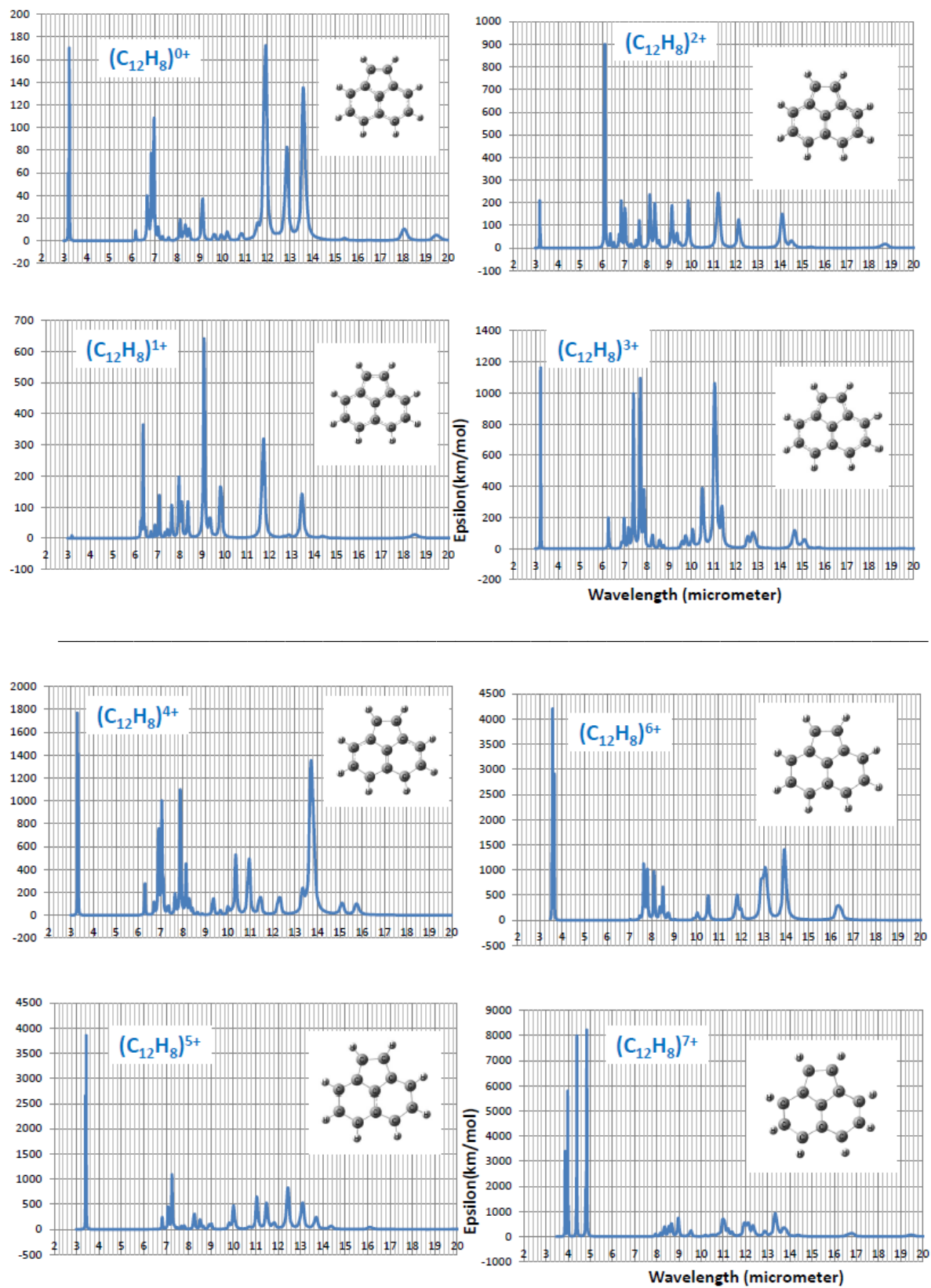

Figure 3, Calculated infrared spectrum of $(C_{12}H_8)^{n+}$ (n=0~7)

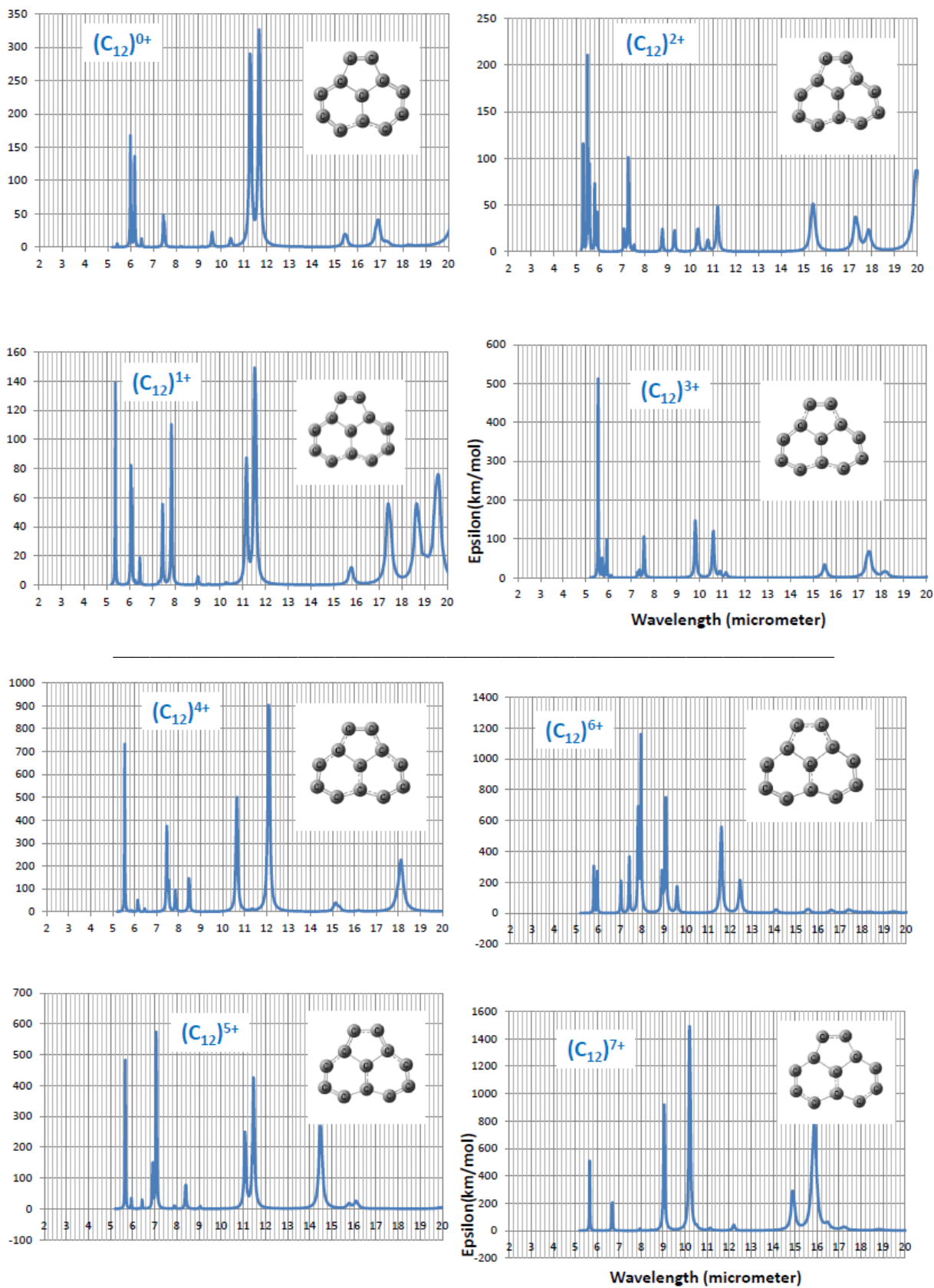

Figure 4, Calculated infrared spectrum of pure carbon molecules of $(C_{12})^{n+}$ (n=0~7)

## 5, HYDROCARBON CONTAINED STARS

### 5.1 WW Vul (WW Vulpeculae, HD344361)

 Observed IR of Herbig Ae star WW Vul (other name: WW Vulpeculae, HD344361) was illustrated in Figure 5 (Acke 2010) by black curve. For comparison, red curve of HD169142 is overlapped as ubiquitously observed typical IR. WW Vul is a famous star exhibiting long-term non-periodic light variability (Udovichenko 2012). Distance is 540pc from earth (Catalogue of Herbig Ae/Be stars, 2018). We can see major bands at 8.2, 9.2, 10.0, and 11.2-11.6 micrometer. Calculated IR of $(C_{12}H_8)^{2+}$ [marked by green curve] and $(C_{12}H_8)^{1+}$ [blue one] can satisfy these bands as shown in Figure 5. It should be noted that by a suitable combination of those two calculated results we can reproduce major band strength. Moreover, detailed bands of sub peaks could be reproduced fairly well as like 12.1, 13.6, and 14.1 micrometer.

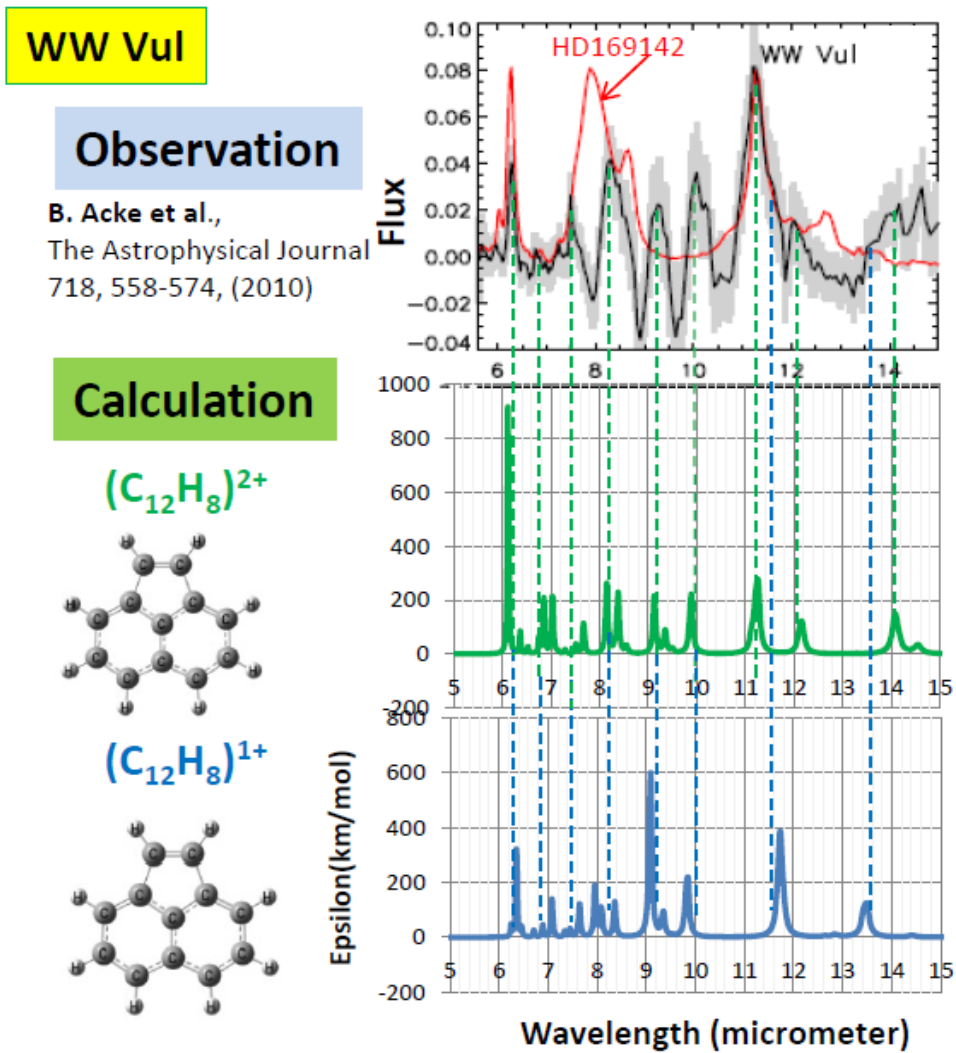

Figure 5, Reproduction of Infrared spectrum of Herbig Ae star WW Vul by a mixture of calculated one of $(C_{12}H_8)^{2+}$ and $(C_{12}H_8)^{1+}$.

## 5.2 HD145263

HD145263 is a Vega-like star with estimated age of 8~10 mega year. Observed spectrum is shown in upper column of Figure 6 by a black curve. This spectrum was also reproduced well by a mixture of calculated one of $(C_{12}H_8)^{2+}$ [green mark] and $(C_{12}H_8)^{1+}$ [blue]. Major bands at 8.4, 9.2, 9.8, and 11.2-11.6 micrometer were reproduced very well, also sub bands fairly well. It remains a difference at 6.2 micrometer where we can see a large peak by calculation, but none for observation.

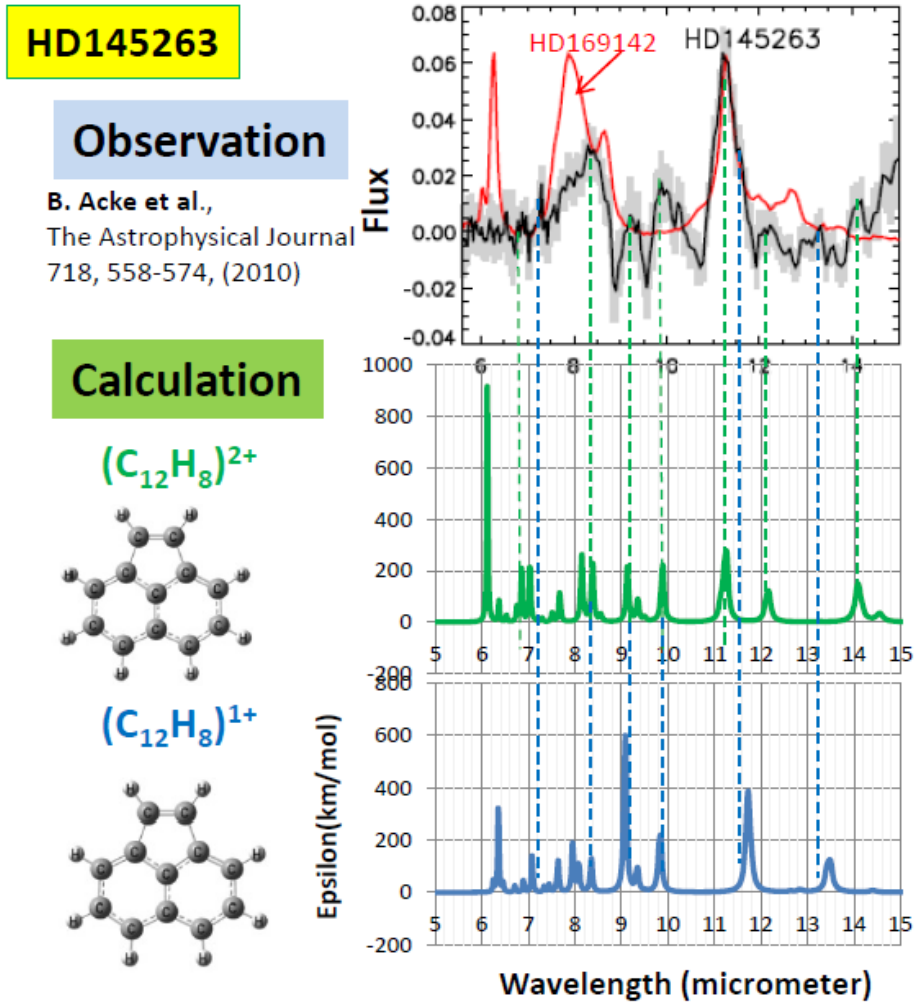

Figure 6, Reproduction of observed infrared spectrum of Herbig Ae star HD145263 by a combination of calculated $(C_{12}H_8)^{2+}$ and $(C_{12}H_8)^{1+}$.

### 5.3 HD37357 ( KMS 27)

HD37357 (other name: KMS 27) is 450pc away star from the earth. Spectrum is shown in Figure 7, which is once previously analyzed by a mixed carrier molecules with $(C_{12}H_8)^{2+}$ and $(C_{12}H_8)^{3+}$ (Ota 2017e).  We analyzed it again by considering $(C_{12}H_8)^{1+}$ [blue curve]. By such an addition, we could reproduce strong feature at 9.3 and 10.0 micrometer, also detailed sub peaks at 11.7 and 13.6 micrometer. Totally, observed 12 bands could be explained by those three molecules mixture model.

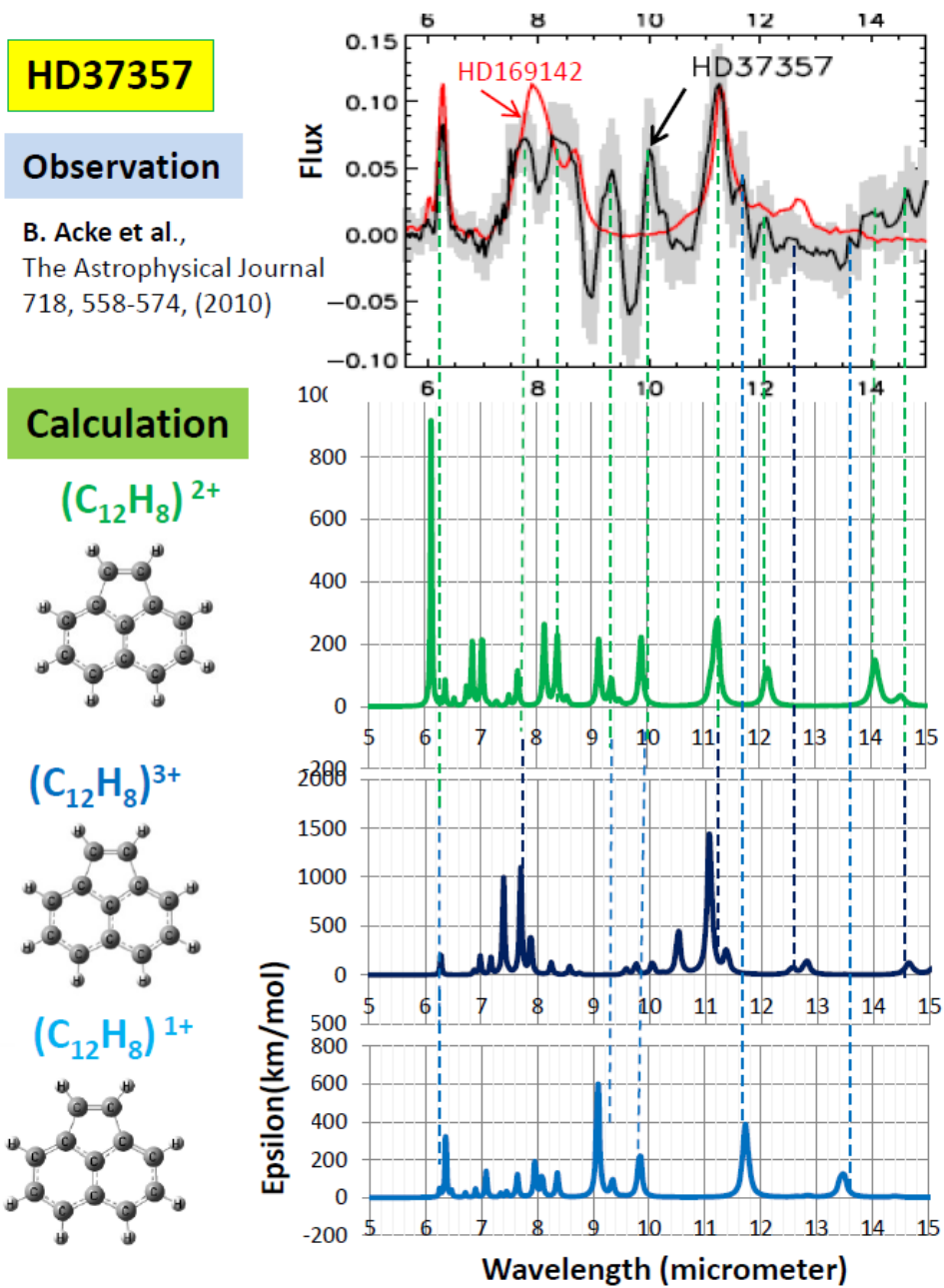

Figure 7, Infrared spectrum reproduction of Herbig Ae star HD37357 by $(C_{12}H_8)^{2+}$, $(C_{12}H_8)^{3+}$ and $(C_{12}H_8)^{1+}$.

## 6, PURE CARBON CONTAINED STARS

Polycyclic pure carbon $(C_{12})^{n+}$ plays an important role to reproduce detailed feature of Herbig Ae stars' infrared spectrum.

### 6.1 HD37258 (V586 Ori)

HD37258 (other name: V586 Ori) is known to have shell like core (Corbally 1991), which is currently supposed to be protoplanetary disk. Major spectrum was explained by di-cation $(C_{12}H_8)^{2+}$ as shown in Figure 8 by green curve . By an addition of pure carbon molecule $(C_{12})^{2+}$ [red broken mark], main band strength at 9.3, and 11.3 micrometer were intensified, shoulder structure at 8.6 and 10.9 micrometer could be supported, and sub peaks at 7.4 and 10.2 micrometer were reproduced well. Moreover, by an addition of neutral $(C_{12}H_8)$, we can explain longer wavelength bands at 12.0, 12.9, and 13.6 micrometer.

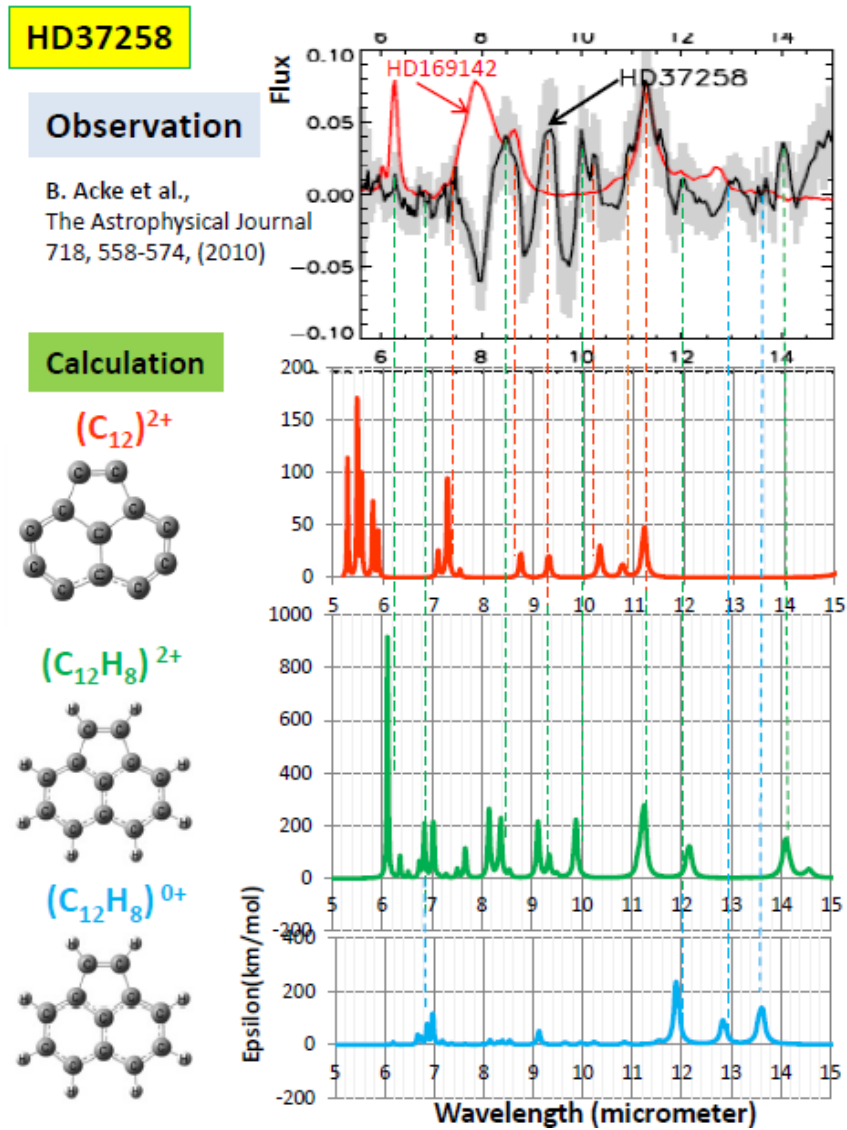

Figure 8, Infrared spectrum reproduction of Herbig Ae star HD37258 by a mixture of pure carbon $(C_{12})^{2+}$, hydrocarbon $(C_{12}H_8)^{2+}$, and $(C_{12}H_8)^{0+}$.

## 6.2 HD38120

HD38120 is 4 times heavier star than our sun with 422 pc distance (Catalogue of Herbig Ae/Be stars 2016). IR spectrum is very complex as shown in Figure 9. Such complexity could be solved by considering calculated spectra both of pure carbon $(C_{12})^{2+}$ [red broken mark] and hydrocarbon $(C_{12}H_8)^{2+}$ [green one]. Major bands at 7.2, 8.5, 9.3, 10.4, 11.3 micrometer were reproduced well by $(C_{12})^{2+}$. Other major bands at 6.2, 8.2, 9.9, 11.3, 12.2, and 14.0 micrometer could be reproduced by $(C_{12}H_8)^{2+}$. Additionally, tri-cation $(C_{12}H_8)^{3+}$ [dark blue] supported sub bands at 7.6, 11.0, 12.7, and 14.6 micrometer bands.

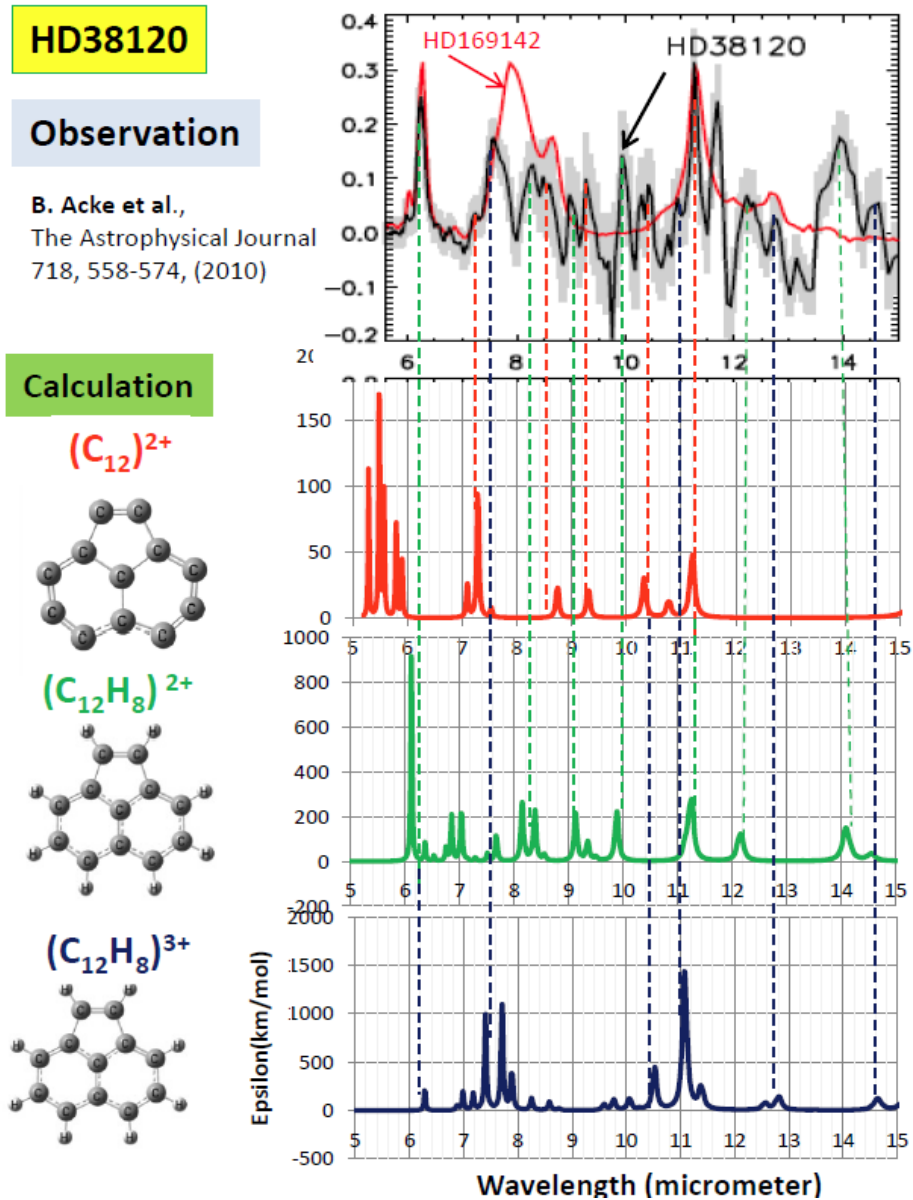

Figure 9, Infrared spectrum of Herbig Ae star HD38120 was well reproduced by a mixture of pure carbon $(C_{12})^{2+}$, hydrocarbon $(C_{12}H_8)^{2+}$, and $(C_{12}H_8)^{3+}$.

## 7, VARIOUS SIZE MOLECULES

Many Herbig Ae stars show ubiquitous IR features as illustrated in Figure 10. Red curve is a typical example by HD169142, which show featured bands at 6.2, 7.8, 8.6, 11.3 micrometer. Whereas, HD142527 presents complex bands as marked by black, where modulates basic ubiquitous pattern by sub bands at 6.9, 7.3, 9.1, 10.0, 12.1, 12.6, and 13.7 micrometer. Our previous studies show that ubiquitous pattern would be explained by coronene and circumcoronene related molecules (Ota 2017a, 2017d, 2017e, 2018). Here again, we analyzed HD142527 by void-induced coronene $(C_{23}H_{12})^{2+}$ [marked by dark green] and void-induced circumcoronene $(C_{53}H_{18})^{1+}$ [light green]. However, in order to explain detailed sub bands, we need third carrier molecule. By applying calculated spectrum in Figure 3, we could find the right one to be $(C_{12}H_8)^{2+}$ [green]. It should be noted that in order to explain total view of carrier molecules around Herbig Ae stars, we need various size molecule candidates.

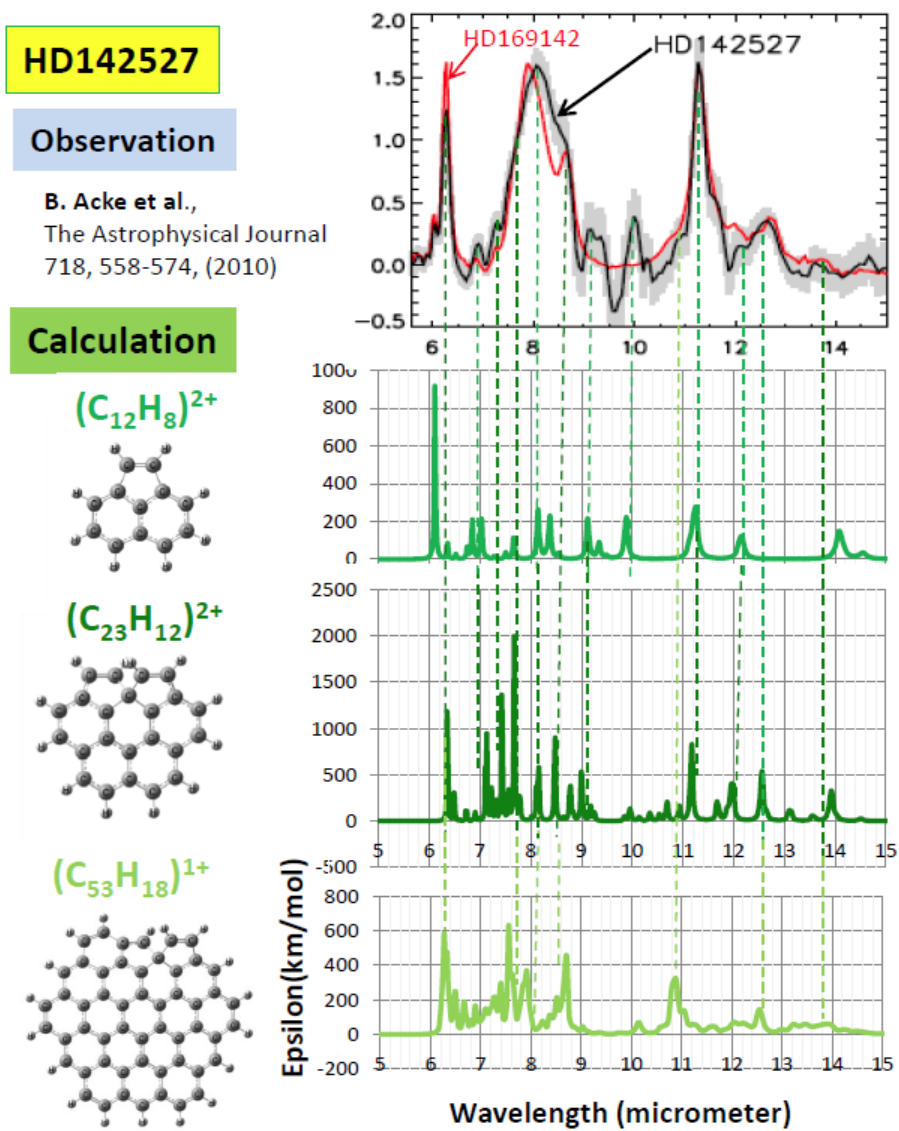

Figure 10, Infrared spectrum reproduction of Herbig Ae star HD142527 by hydrocarbon $(C_{12}H_8)^{2+}$, $(C_{23}H_{12})^{2+}$ and $(C_{53}H_{18})^{1+}$. In order to explain total view of carrier molecules around Herbig Ae stars, we need various size molecule candidates.

## 7, CONCLUSION

Astrochemical evolution step of polycyclic aromatic hydrocarbon (PAH) around a Herbig Ae young star was analyzed using the first principles quantum chemical calculation. For simplicity, acenaphthylene ($C_{12}H_8$) with hydrocarbon one pentagon combined with two hexagons was studied and applied for carrier molecule identification.

(1) In a protoplanetary disk around a just born star, PAH's would be attacked by high energy photon from the central star and ionized to be high order cation as like $(C_{12}H_8)^{n+}$ .

(2) Calculation shows that from n=0 to 6, molecules keeps its polycyclic hydrocarbon configuration. Whereas, ionization step at n=7, there occur dehydrogenation resulting polycyclic pure carbon ($C_{12}$).

(3) Such pure carbon molecule would be illuminated again by photons. At a stage of eighth ionization $(C_{12})^{8+}$, there occur decomposition to aliphatic carbon chains, $C_9$, $C_2$, and mono carbon $C_1$.

(4) Carrier molecules of Herbig Ae star WW Vul (also HD145263) were identified by a combination of $(C_{12}H_8)^{2+}$ and $(C_{12}H_8)^{1+}$. Complex spectrum of HD37357 could be explained by $(C_{12}H_8)^{2+}$, $(C_{12}H_8)^{3+}$, and $(C_{12}H_8)^{1+}$.

(5) Pure carbon molecules play an important role. IR of HD37258 was analyzed by a mixture of pure carbon $(C_{12})^{2+}$, hydrocarbon $(C_{12}H_8)^{2+}$ and neutral $(C_{12}H_8)$. Also, complex spectrum of HD38120 was analyzed by $(C_{12})^{2+}$, $(C_{12}H_8)^{2+}$ and $(C_{12}H_8)^{3+}$.

Acenaphthylene related molecules are just a typical example. We should apply various size molecules for studying astrochemical evolution mechanism around a new born star.

## ACKNOWLEDGEMENT

I would like to say great thanks to Prof. Takashi Onaka and Prof. Itsuki Sakon, University of Tokyo, for very kind discussion on infrared astronomy.

Author profile
   Norio Ota PhD.
    Senior Professor, University of Tsukuba, Japan
    Material Science, Optical data storage devices

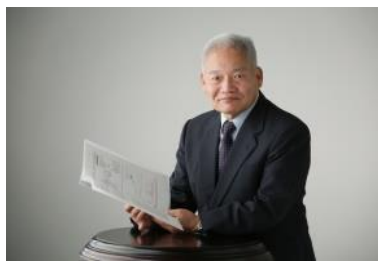

**Submit to arXiv.org. ,      March          , 2018   by Norio Ota**